\documentclass[twocolumn,showpacs,preprintnumbers,amsmath,amssymb]{revtex4}

\usepackage{graphicx}% Include figure files
\usepackage{dcolumn}% Align table columns on decimal point
\usepackage{bm}% bold math

\begin{document}

\title{A remark on the local density approximation with the gradient corrections and the X$_\alpha$ method}

\author{Y.-B. Xie}\email{xyb@ustc.edu.cn}

\author{B.-H. Wang}

\author{W.-X. Wang}

\affiliation{%
Department of Modern Physics, University of Science and Technology
of China, Hefei, 230026, P.R. China }%

\begin{abstract}
We report that the solids with narrow valence bands cannot be
described by the local density approximation with the gradient
corrections in the density functional theory as well as the
X$_\alpha$ method. In particular, in the case of completely filled
valence bands, the work function is significantly underestimated by
these methods for such types of solids. Also, we figured out that
these deficiencies cannot be cured by the
self-interaction-corrected-local-density-approximation method.
\end{abstract}

\pacs{71.15.Mb, 31.15.Ew}

\maketitle

\section{Introduction}
The density functional theory \cite{DFT,LDA} with the generalized
gradient approximations \cite{GGA1,GGA2} (GGA) and the X$_\alpha$
theory \cite{Xa,Xa23} have been quite successful in the calculation
of various properties in many systems\cite{Levine}, such as the bond
length, the bond angle, various properties related to the charge
density and the vibration frequencies of many moleculars and solids.
Moreover, the binding energies of many moleculars and the cohesive
energies of many solids can be as well roughly evaluated by the GGA
and the X$_\alpha$ theory.

Despite those successes, it is also well known\cite{Kunz} that the
single particle electronic structure cannot be very well described
by the GGA in many solids, since the GGA tends to underestimate the
valence band width and the band gap.  This is not difficult to
understand because even in the electron gas model, the exchange
energies for the states far below or far above the Fermi surface are
significantly underestimated or overestimated, respectively.
Nevertheless, the exchange energy for the states at the Fermi level
is correctly given by the GGA for the electron gas case. Therefore,
the properties related to the electronic structure at the Fermi
surface, for instance, the work function, has been considered
correctly explained by the GGA for many solids.

On the other hand, the eigenvalues of the X$_\alpha$ single electron
equation cannot be used to estimate the ionization energies of atoms
and small moleculars. Fortunately, this problem can be cured by
introducing the so called transition state method\cite{Xa}.

Let us consider the solids in which the valence electronic states
deviates from the plane waves significantly.  The electronic
properties of this type of solids cannot be described by the
electron gas model.  In contrast, some characters of the individual
atom or small molecular are still kept in such solids.  But since
the single particle states are quite extended in crystalline solids,
the transition state method will no longer be helpful.  Hence, a
natural question is addressed: can the GGA or the X$_\alpha$ method
correctly describe the single particle states near the Fermi surface
for such type of solids?  We figured out that the answer to this
question, unfortunately, is negative from both the academic and
practical point of view.  Our paper will provide enough evidence to
support this conclusion.

In this paper, we study a system composed of spin polarized hydrogen
atoms being located on a cubic lattice.  The reason to choose such
system is for simplicity.  Since the spins of all electrons are
parallel, the correlation energy is insignificant, and actually is
zero in the case of large lattice constant. The exchange energy can
be treated exactly by the Hartree Fock method. Also the exchange
functional in the GGA and the X$_\alpha$ method is simple and the
correlation functional is negligibly small in this system.

In section III, we consider the large $a$ case in which the overlap
of the localized hydrogen electron wavefunctions on different sites
is negligible.  We rigorously show that in GGA or in X$_\alpha$
method, the ionization energy in such case is given by the
eigenvalue of the single particle equation for one hydrogen atom
which is roughly\cite{units} 0.29 and deviates from the exact
ionization energy $0.5-37/a^4$ considerably (Note that here the term
$37/a^4$ is the hole polarization term.). This difference, as we can
see, is present not only in the spin polarized hydrogen solids, but
also in many other systems, where atoms are well separated and the
single particle wavefunctions are extended. Notice here that the
total bulk energy can be still correctly given by the GGA or the
X$_\alpha$ theory.

In section IV, we study the finite $a$ case in which the overlap of
the localized wavefunctions on different sites cannot be neglected.
In this case, the single particle wavefunctions must be extended.
Therefore, similar to the reported results in the last paragraph, we
expect that the work function is still significantly underestimated
by the GGA or the X$_\alpha$ method when $a$ is not small.  This
expectation is indeed correct up to $a=5$ at which the band width is
nearly $0.25$. When $a$ is sufficiently small, the single particle
states can be characterized by the plane wave and the work function
should be correctly predicted by the GGA or the X$_\alpha$ theory.

Finally in section V, we summarize our results.

\section{The model}

Let us consider $N$ hydrogen atoms located on a cubic lattice with
the lattice constant $a$.  For simplicity, we assume that the spins
of all electrons are up.  In this paper, we study the work function
for the $N$ electrons system, i.e., the ground state energy
difference between the $N$ electrons system and the $N-1$ electrons
system with fixed $a$.

We investigate this system using the Hartree Fock method and the GGA
or the X$_\alpha$ method.  The correlation energy is not significant
because the spins of the electrons are parallel. Actually, the
contribution of the correlation energy to the work function is
estimated to be $-37/a^4$ when $a$ is large. Hence, the Hartree Fock
method is accurate to predict the work function. On the other hand,
the GGA is also quite simple in this case because only the exchange
functional is significant whileas the correlation functional is
small (In this paper, we shall ignore the correlation functional
because it is absent for a single hydrogen atom.). We focus on the
error of the work function predicted by the GGA or the X$_\alpha$
method.

When $a$ is finite, the occupied single electron states are
described by the Bloch state $|{\bf k}>=\int d^3{\bf r} u_{\bf
k}({\bf r})|{\bf r}>$ with $|k_x|,|k_y|,|k_z|<\pi/a$.  In the
framework of Hartree Fock theory, $u_{\bf k}=u^{{\rm HF}}_{{\bf k}}$
satisfy the Hartree Fock equation
\begin{eqnarray}
&&(-{1\over 2}\nabla^2+U({\bf r})+\sum_{{\bf p}}\int d^3{\bf r}'
{u^{{\rm HF}*}_{{\bf p}}({\bf r'})u^{{\rm HF}}_{{\bf p}}({\bf
r}')\over |{\bf r-r'}|}) u^{{\rm HF}}_{{\bf
k}}({\bf r})\nonumber\\
&-&\sum_{{\bf p}}\int d^3{\bf r}' {u^{{\rm HF}*}_{{\bf p}}({\bf
r'})u^{{\rm HF}}_{{\bf p}}({\bf r})u^{{\rm HF}}_{{\bf k}}({\bf
r}')\over |{\bf r-r'}|}\nonumber\\
&&=\epsilon_{{\rm HF}}({\bf k})u^{{\rm HF}}_{{\bf k}}({\bf r}),\\
\end{eqnarray}
where
\begin{equation}
U({\bf r})=-\sum_{{\bf R}_i}{1\over |{\bf r-R}_i|}
\end{equation}
with ${\bf R}_i$ representing the lattice points. The Hartree Fock
orbital energy can be also written as
\begin{equation}
\epsilon_{{\rm HF}}({\bf k})=<{\bf k}|T+U|{\bf k}>+\sum_{{\bf
p}}({\bf pp}|{\bf kk})-({\bf pk}|{\bf kp})
\end{equation}
with $T=-\nabla^2/2$ and \begin{equation} ({\bf k_1k_2}|{\bf
k_3k_4})=\int d^3{\bf r}\int d^3{\bf r}'{u_{{\bf k}_1}^*({\bf
r})u_{{\bf k}_2}({\bf r})u_{{\bf k}_3}^*({\bf r}')u_{{\bf k}_4}({\bf
r}')\over |{\bf r-r}'|}. \end{equation}
 Because the Bloch states
$|{\bf k}>$ are extended, the Hartree Fock wavefunctions $u_{{\bf
p}}({\bf r})$ will be not modified when one electron in $|{\bf k}>$
is removed. According to the Koopman's theorem, the ionization
potential associated with the state $|{\bf k}>$ is simply given by
$-\epsilon_{{\rm HF}}({\bf k})$.  By adding the correction from the
correlation energy, one obtains the exact ionization potential.
Since the state at the $X$ point ${\bf k}=({\pi/a,\pi/a,\pi/a})$ is
at the top of the electronic band, the work function is
approximately given by $-\epsilon_{{\rm HF}}(X)$.

In the framework of GGA, $u_{{\bf k}}=u^{{\rm GGA}}_{{\bf k}}$ which
is determined by the Kohn-Sham equation
\begin{eqnarray}
&&(-{1\over 2}\nabla^2+U({\bf r})+\int d^3{\bf r}' n({\bf r})/|{\bf
r-r'}|) u^{{\rm GGA}}_{{\bf k}}({\bf r})\nonumber\\
&+&v_{{\rm xc}}(n({\bf r})) u^{{\rm GGA}}_{{\bf k}}({\bf r})
=\epsilon_{{\rm GGA}}({\bf k})u^{{\rm GGA}}_{{\bf k}}({\bf r})
\end{eqnarray}
with the exchange-correlation potential given by
\begin{equation}
v_{{\rm xc}}(n({\bf r}))={\delta E_{{\rm xc}}[n]\over\delta n({\bf
r})},
\end{equation}
where $E_{{\rm xc}}[n]$ is the exchange-correlation functional and
\begin{equation}
n({\bf r})=\sum_{{\bf p}} u^{{\rm GGA}^*}_{{\bf p}}({\bf r}) u^{{\rm
GGA}}_{{\bf p}}({\bf r})
\end{equation}
is the total electron density. The GGA orbital energy can be also
written as
\begin{eqnarray}
&&\epsilon_{{\rm GGA}}({\bf k}) =\int d^3 {\bf r}u^{{\rm
GGA}*}_{{\bf k}}({\bf r})(-{1\over 2}\nabla^2)u^{{\rm GGA}}_{{\bf
k}}({\bf r})\nonumber\\
 +&&[U({\bf r})+\int d^3{\bf r'} {n({\bf r'})\over |{\bf r'-r}|}
+v_{{\rm xc}}(n({\bf r}))]n_{{\bf k}}({\bf r})
\end{eqnarray}
with
\begin{equation}
n_{{\bf k}}({\bf r})=u^{{\rm GGA}*}_{{\bf k}}({\bf r}) u^{{\rm
GGA}}_{{\bf k}}({\bf r}).
\end{equation}

 The total electronic energy of this system is
given by
\begin{eqnarray}
E&=&\sum_{{\bf k}}\int d^3{\bf r}u^{{\rm GGA}*}_{{\bf k}}({\bf
r})(-{1\over 2}\nabla^2)u^{{\rm GGA}}_{{\bf k}}({\bf r})
\nonumber\\
&+&\int d^3{\bf r} U({\bf r})n({\bf r})+{1\over 2}\int d^3{\bf
r}d^3{\bf r'} {n({\bf r})n({\bf r'})\over |{\bf r-r'}|}
\nonumber\\
&+&E_{{\rm xc}}[n]+U_{NN},
\end{eqnarray}
where \begin{equation} U_{NN}=\sum_{i<j} {1\over |{\bf R}_i-{\bf
R}_j|}. \end{equation} Next, we consider the $N-1$ electrons system
with the $|{\bf k}=X>$ state unoccupied. Since $|X>$ is extended,
the total electron density $n'({\bf r})$ for the $N-1$ electrons
system is approximately equal to $n({\bf r})$ in the large $N$
limit. Thus the GGA solution $u^{{\rm GGA}}_{{\bf k}}({\bf r})$ is
not modified and we self-consistently confirms that $n'\approx n$.
The total ground electron energy for the $N-1$ system is given by
\begin{eqnarray}
E'&=&\sum'_{{\bf k}}\int d^3{\bf r}u^{{\rm GGA}*}_{{\bf k}}({\bf
r})(-{1\over 2}\nabla^2)u^{{\rm GGA}}_{{\bf k}}({\bf r})
\nonumber\\
&+&\int d^3{\bf r} U({\bf r})n'({\bf r})+{1\over 2}\int d^3{\bf
r}d^3{\bf r'} {n'({\bf r})n'({\bf r'})\over |{\bf r-r'}|}
\nonumber\\
&+&E_{{\rm xc}}[n']+U_{NN},
\end{eqnarray}
where $\sum'_{{\bf k}}$ means the summation over all ${\bf k}$
states except the $X$ state and $n'({\bf r})=n({\bf r})-n_X({\bf
r})$.  Since $n_X\ll n$ in the large $N$ limit, we obtain the work
function
\begin{equation}
W=E'-E=-\epsilon^{{\rm GGA}}(X)
\end{equation}
which is just the orbital energy of the state $|X>$.  Eq.(14) can be
also obtained by the transition state method generally used in the
X$_\alpha$ method.  It is worthwhile to point out that for finite
system or when the hole state is localized, Eq.(14) is no longer
valid.

From many calculations for atoms and small moleculars, we know that
the orbital energy obtained in the GGA or the X$\alpha$ method is
significantly lower than the exact ionization potential. Therefore,
one suspects that $W$ is significantly underestimated by Eq.(14) in
solids in which the valence electrons cannot be described by the
plane wave states.  In the following sections, this suspection has
been proved to be true.

\section{Non overlapping case}

In this section, we study the large $a$ case in which the overlap
between the atomic wavefunctions on different sites is negligible.
In this case, one can immediately obtain the solution in the Hartree
Fock theory and in the GGA or the X$_\alpha$ method. Actually in the
Hartree Fock theory, we have
\begin{equation}
u^{{\rm HF}}_{{\bf k}}({\bf r})={1\over\sqrt{N}}\sum_{{\bf R}_i}
e^{i{\bf k}\cdot{\bf R}_i}\phi^{{\rm HF}}_0({\bf r-R}_i)
\end{equation}
with $\phi^{{\rm HF}}_0({\bf r})$ being the exact ground state
wavefunction for a single hydrogen atom located at ${\bf R}=0$, and
\begin{equation}
\epsilon^{{\rm HF}}({\bf k})=-0.5.
\end{equation}
Therefore, the work function predicted by the Hartree Fock
approximation is $W^{{\rm HF}}=0.5$.  There is a correction to $W$
from the correlation energy in the $N-1$ electrons system
corresponding to the hole polarization term.  The magnitude of this
correction is $-37/a^4$ (We have ignore the Van der Waals
interaction term which is very small.).

On the other hand, in the GGA or the X$_\alpha$ method, we have
\begin{equation}
u^{{\rm GGA}}_{{\bf k}}({\bf r})={1\over\sqrt{N}}\sum_{{\bf R}_i}
e^{i{\bf k}\cdot{\bf R}_i}\phi^{{\rm GGA}}_0({\bf r-R}_i)
\end{equation}
with $\phi^{{\rm GGA}}_0({\bf r})$ being the ground state
wavefunction in the GGA or the X$_\alpha$ method for a single
hydrogen atom located at ${\bf R}=0$ and $\phi^{{\rm GGA}}_0({\bf
r})$ satisfying the following equation
\begin{eqnarray}
&&(-{1\over 2}\nabla^2-{1\over r}+\int d^3 {\bf r}' {n_0({\bf
r}')\over |{\bf r-r}'|}\nonumber\\
&&+v_{{\rm xc}}(n_0^{{\rm GGA}}({\bf r})))\phi^{{\rm GGA}}_0({\bf
r})=\epsilon_0^{{\rm GGA}}\phi^{{\rm GGA}}_0({\bf r}),
\end{eqnarray}
where
\begin{equation}
n_0^{{\rm GGA}}({\bf r})=|\phi^{{\rm GGA}}_0({\bf r})|^2
\end{equation}
with
\begin{eqnarray}
&&\epsilon^{{\rm GGA}}({\bf k})=\epsilon^{{\rm GGA}}_0 = \int d^3
{\bf r}\phi^{{\rm GGA}*}_0({\bf r})(-{1\over 2}\nabla^2)\phi^{{\rm
GGA}}_0({\bf r})\nonumber\\
& +&[-{1\over r}+\int d^3{\bf r'} {n_0^{{\rm GGA}}({\bf r'})\over
|{\bf r'-r}|} +v_{{\rm xc}}(n_0^{{\rm GGA}}({\bf r}))]n_0^{{\rm
GGA}}({\bf r}).
\end{eqnarray}
The above equation indicates that the work function $W$ is indeed
given by the orbital energy of the single atom in the GGA or the
X$_\alpha$ method in this case.  However, it is well known that the
orbital energy of a single atom in the GGA or the X$_\alpha$ method
is much less in magnitude than the corresponding ionization
potential. For hydrogen atom, $\epsilon^{{\rm GGA}}_0$ is found to
be $-0.29$ in X$_\alpha$ method, in which $\alpha$ is chosen such
that the self interaction energy is exactly canceled by the exchange
functional $E_{{\rm x}}[n]$ (Note that we have used the exact
wavefunction $\phi^{{\rm HF}}_0({\bf r})$ in Eq.(20) to calculate
$\epsilon^{{\rm GGA}}_0$).  The difference between the work function
given by the Hartree Fock theory and that given by the X$_\alpha$
method is $(00|00)/3\approx 0.208$ with
\begin{equation}
(00|00)=\int d^3{\bf r} d^3{\bf r}'{n^{{\rm HF}}_0({\bf r})n^{{\rm
HF}}_0({\bf r}')\over |{\bf r-r}'|}=0.625.
\end{equation}
 It should be
pointed out that the difference between the exact $\epsilon^{{\rm
GGA}}_0$ evaluated by $\phi^{{\rm GGA}}_0({\bf r})$ and the above
value is insignificant\cite{GGA1}. In GGA, the magnitude of
$\epsilon^{{\rm GGA}}_0$ is even less than 0.29. Thus, the work
function is remarkably underestimated in this case by the GGA or the
X$_\alpha$ method.

We need to point out that the above conclusion is not only
applicable to the hypothetic solid in which the spin polarized
hydrogen atoms are located on a cubic lattice, but also applicable
to the real crystalline solids, with spin up and spin down
electrons, satisfying the following two conditions (i) The energy
bands are either completely filled or completely empty. (ii) The
lattice constant is sufficiently large such that the overlap of the
wavefunctions between different lattice sites is negligible.  This
conclusion can be derived using the exactly same way as for our
hypothetic solid. Many real solids for instance, the solids formed
by Ne, Ar, Kr, Xe atoms at low temperatures and even ice (Note that
there is polarization induced by the dipole moment by the nearby
water molecular.  But this effect is very small.) may satisfy the
above two conditions.  It should be noticed that for instance for
the Ne solid, the work function is obtained by the Hartree Fock
orbital energy together with the corrections from the electron
correlation effect as well as the lattice relaxation effect (i.e.,
the polaron effect). The magnitude of those corrections should be
less than 0.1.  However, the difference of the orbital energy
between the HF theory and the GGA is estimated as 0.4. So the
underestimation of the work function for the solid Ne given by the
GGA is expected to be quite significant (about 0.3).

Although the ionization potential is remarkably underestimated in
the non-overlapping case, the total electronic energy $E$ is still
correctly given by the GGA or the X$_\alpha$ method when the band is
completely filled.  Since
\begin{equation}
n^{{\rm GGA}}({\bf r})\approx\sum_{{\bf R}_i}n^{{\rm GGA}}_0({\bf
r-R}_i)
\end{equation}
and
\begin{equation}
n_0^{{\rm GGA}}({\bf r-R}_i)n_0^{{\rm GGA}}({\bf r-R}_j)\approx 0
\end{equation}
when ${\bf R}_i\ne{\bf R}_j$, we have
\begin{eqnarray}
&&E=N\int d^3 {\bf r}\phi^{{\rm GGA}*}_0({\bf r})(-{1\over
2}\nabla^2)\phi^{{\rm GGA}}_0({\bf r})\nonumber\\
& +&[-{1\over r}+{1\over 2}\int d^3{\bf r'} {n_0^{{\rm GGA}}({\bf
r'})\over |{\bf r'-r}|}]n_0^{{\rm GGA}}({\bf r})+E_{{\rm
xc}}[n_0^{{\rm GGA}}] \nonumber\\
&& \approx -0.5N.
\end{eqnarray}

Since the bandwidth is vanishing in the non-overlapping case, it may
be interesting to consider the localized hole state.  In this case,
the ionization potential equals $0.5-37/a^4$ and is correctly given
by the GGA or the X$_\alpha$ method.  Unfortunately, the GGA
predicts that the energy of the extended hole state is lower than
the energy of the localized hole state.

It may be pedagogic to discuss the case that the band is only
partially filled.  Here, for simplicity, we shall only consider the
infinite $a$ case.  Suppose that there are $Z\leq N$ spin polarized
electrons in the system.  Assuming that the single particle states
are all extended over the whole lattice.  In the X$_\alpha$ theory,
the total electronic energy is found to be
\begin{equation}
E(Z,N)\approx -{1\over 2}Z(1-{(00|00)\over 2}[{Z\over N}-({Z\over
N})^{1\over 3}])^2.
\end{equation}
(Note that this equation is valid only when $a=\infty$.  When $a$ is
large but finite, there is an additional term $C/a$ for $E(Z,N)$
when $Z\ne N$.) In Eq.(2), we have used $\phi^{{\rm HF}}_0(\xi{\bf
r})$ in the construction of Bloch states, where $\xi$ is a
variational scaling parameter which is used to minimize the total
X$_\alpha$ energy. Therefore, the total electronic energy is
significantly overestimated when the band is only partially filled.
Interestingly, when $Z\ll N$, one may even obtain a lower value for
$E(Z,N)$ by assuming all single particle states are extended only
over five lattice sites. According to Eq.(25), one may obtain the
ionization potential for $Z$ electrons
\begin{eqnarray}
&&I(Z,N)=E(Z-1,N)-E(Z,N)\nonumber\\
 &=&{1\over 2}[1-{(00|00)\over 2}[{Z\over N}-({Z\over
 N})^{1/3}]]\nonumber \\
&& [1-{(00|00)\over 2}[3{Z\over N}-{5\over 3}({Z\over N})^{1/3}]].
\end{eqnarray}
Notice that when the band is half filled $Z=N/2$, we have $I=0.516$
which is quite close to the exact value $0.5$

It may be necessary to mention that this deficiency in the GGA or
the X$_\alpha$ method is also found \cite{SIE1,SIE2,SIE3} in some
molecular systems at the dissociation limit, such as H$_2^+$,
F$_2^+$, NaCl, et.al. Essentially, both failures of the GGA or the
X$_\alpha$ method for the solids at large $a$ and for moleculars at
the dissociation limit are caused by the same reason.  The authors
in Ref. \cite{SIE1,SIE2,SIE3} attributed this type of failure to the
so called self-interaction error.  However, we deem that this
viewpoint is inappropriate because the self-interaction error is
zero in our system, yet the magnitude of the total electronic energy
is still significantly overestimated by the X$_\alpha$ method.  This
fact also suggests that the self-interaction-corrected local density
approximation method \cite{SIC1, SIC2} is not helpful to improve the
result.

We notice as well that the overestimation of the magnitude of the
electronic energy by the X$_\alpha$ method is more pronounced in our
system than in diatomic molecular cases, because the electronic
states are much more extended in our case. For instance, consider
the ionization potential of two spin polarized hydrogen atoms at the
fixed separation distance $a$.  When $a$ is large and the atomic
wavefunction overlap between two hydrogen atoms is negligible small,
we can obtain that the error of the ionization potential given by
the X$_\alpha$ method
\begin{eqnarray}
&&\delta I=(00|00)[({1\over 2})^{4/3}-{1\over 4}]-{1\over
4a}\nonumber\\
&&\approx 0.092-{1\over 4a},
\end{eqnarray}
where the polarization term has been neglected because it is
proportional to $1/a^4$.  This error is significantly lower than
that in solid case.  For instance, when $a=5$, this number is only
0.04 which is much less than 0.21 in the solid case. This fact
explains why the GGA or the X$_\alpha$ method are still successful
for small molecular near equilibrium geometry.

\section{Finite overlap case}
In the last section, we have discussed the large $a$ case in which
the overlap of the wavefunctions on different sites is negligible.
However, the GGA or the X$_\alpha$ method are usually applied to
real solids of finite bandwidth.  When $a$ is sufficiently small,
the electrons in our system can be described by the plane wave.
Therefore, one may speculate that the GGA becomes valid for not
large $a$ case in our system.  On the other hand, for large $a$, the
atomic nature still remains and it can be expected that the error of
the work function predicted by the GGA or the X$_\alpha$ method is
still significant even when the bandwidth is finite.  So it is
highly desirable to see how this error decreases when $a$ decreases.

When $a$ is finite and the bandwidth is not zero, it is not easy to
calculate the electronic energy accurately even within the framework
of the HF theory or the GGA.  Instead, we shall take several
approximations which can greatly simplify the calculations.

At first, we assume $u^{{\rm HF}}_{{\bf k}}({\bf r})\approx u^{{\rm
GGA}}_{{\bf k}}({\bf r})$ and calculate the difference of the work
function predicted by the Hartree Fock theory and the GGA or the
X$_\alpha$ method.  When the hole state is extended, straightforward
calculation yields
\begin{equation}
\Delta W=W_{{\rm HF}}-W_{{\rm GGA}} =\Delta_1-\Delta_2
\end{equation}
with
\begin{equation} \Delta_1=\sum_{{\bf k}}({\bf k}X|X{\bf
k}), \end{equation} \begin{equation} \Delta_2=\int d^3{\bf r}
v_{{\rm xc}}(n({\bf r}))n_{{\rm X}}({\bf r}),
\end{equation}
where $n$ and $n_X$ are the total electron density and the electron
density of the state $|X>$, respectively, $W_{{\rm HF}}$ and
$W_{{\rm GGA}}$ are the work function obtained by the Hartree Fock
theory and the GGA, respectively.  Note that this assumption has
been tested in atomic calculations.  The error from this assumption
is found to be insignificant\cite{GGA1}.

Secondly, we assume
\begin{equation}
u_{{\bf k}}({\bf r})={C({\bf k})\over \sqrt{N}}\sum_{{\bf R}_i}
e^{i{\bf k}\cdot{\bf R}_i} \phi^{{\rm HF}}_0({\bf r-R}_i)
\end{equation}
with the normalization constant
\begin{equation}
C({\bf k})={1\over\sqrt{\sum_{{\bf R}_i} S({\bf R}_i) e^{i{\bf
k}\cdot{\bf R}_i}}},
\end{equation}
where the overlap is defined as
\begin{equation}
S({\bf R})=\int d^3{\bf r}\phi^{{\rm HF}}_0({\bf r})\phi^{{\rm
HF}}_0({\bf r-R})
\end{equation}
This assumption is valid in the tight binding limit.  While for
small $a$ and large bandwidth, different $|{\bf k}>$ states are made
of different $\phi$ and this assumption becomes invalid. However,
since both the quantities $<{\bf k}X|X{\bf k}>$ and $v_{{\rm xc}}$
do not depend on $u_{{\bf k}}$ sensitively, we expect that this
assumption is valid at least when $a$ is not small. Under this
assumption, the work function difference $\Delta W$ can be evaluated
numerically.

Thirdly, we shall use the X$_\alpha$ exchange functional with the
parameter $\alpha=\alpha_{{\rm H}}$ selected to cancel the
self-interaction exactly in the atomic limit. So we have
\begin{equation}
E_{{\rm xc}}[n({\bf r})]={20\pi^{1/3}\over 27}\int d^3{\bf
r}n^{4/3}({\bf r}).
\end{equation}
Since $\alpha_{{\rm H}}$ is greater than the corresponding value in
the electron gas model, and furthermore our system will reduce to
the electron gas model in the small $a$ limit, we expect that this
exchange functional will slightly overestimate the exchange energy
in the finite $a$ case.
\begin{table}
\begin{tabular}{|c|c|c|c|c|c|c|}\hline
$a$  & 3.0 & 4.0   & 5.0   & 6.0  &7.0 & $\infty$ \\ \hline
$\Delta_1$ & 0.859 & 0.738 & 0.685 & 0.656 & 0.639 & 0.625
\\ \hline $\Delta_2$ &0.716  & 0.577  & 0.504 & 0.465 & 0.442 &0.417 \\ \hline$\Delta W$
& 0.143 & 0.161 & 0.181 & 0.191& 0.197& 0.208 \\ \hline
\end{tabular}
\caption{The calculated results for $\Delta_1, \Delta_2, \Delta W$
at various values of $a$.}
\end{table}
Table.1 lists the numerical value of $\Delta_1$, $\Delta_2$ and
$\Delta W$ for various values of $a$. From this table, one can see
that when $a\geq 5$,$\Delta W$ significantly deviates from zero
(Note that when $a=3$ and $a=4$, we expect that the second
approximation becomes invalid.  For instance when $a=4$, the
bandwidth is estimated as 0.5 and the electronic states cannot be
described by Eq.(31).). Even after subtracting the correction from
the correlation energy contribution $37/a^4$, the underestimation of
the work function by this X$_\alpha$ method is still about 0.12 at
$a=5$. On the other hand, the band width is about 0.25 when $a=5$.
Therefore, we can conclude that the work function will be
underestimated by the X$_\alpha$ method when (i) the valence bands
are completely filled and (ii) the bandwidth of the valence bands is
not large.

We have also used the generalized gradient approximation \cite{GGA2}
 exchange functional to calculate $\Delta_2$.  In GGA, the exchange
functional is expressed as \cite{GGA2}
\begin{equation}
E_{{\rm xc}}[n]=A\int d^3 {\bf r} n^{4/3} F(t)
\end{equation}
with $A=-(3/4)(6/\pi)^{1/3}$,
\begin{equation}
F(t)=(1+0.021326t+0.0037909t^2+0.000000891t^3)^{1/15}
\end{equation}
and
\begin{equation}
t={(\nabla n)^2\over n^{8/3}}.
\end{equation}
And $\Delta_2$ is evaluated by
\begin{eqnarray}
&&\Delta_2=A\int d^3{\bf r} {4\over 3} n^{1/3} n_{{\rm X}}
F(t)\nonumber\\
&& + F'(t)[{2\nabla n\cdot\nabla n_{{\rm X}}\over n^{4/3}} -{8\over
3} {(\nabla n)^2 n_{{\rm X}}\over n^{7/3}}].
\end{eqnarray}
As expected, $\Delta_2$ evaluated by this formula is indeed slightly
less than that by the X$_\alpha$ method.  For instance, when $a=4$,
$\Delta_2$ is found to be 0.548 in GGA which is slightly less than
the value 0.577 given by X$_\alpha$ method.

\section{Discussion}
In this paper, we have studied the hypothetic system composed of the
hydrogen atoms with spin polarized electrons on a cubic lattice. We
found that for the case of narrow bandwidth and completely filled
band, the work function is significantly underestimated by the GGA
or the X$_\alpha$ method. This is attributed to the significant
lower energy of the extended hole state obtained by the GGA or the
X$_\alpha$ method. Note that this problem cannot be cured by
assuming that the hole state is localized. Actually, the GGA or the
X$_\alpha$ method predicts a significant lower energy for the
extended hole state than for the localized hole state even in the
zero bandwidth limit.

Although we have only calculated a hypothetic system, our results
may lead us to speculate the validness of the GGA or the X$_\alpha$
method in the calculations of the work function of the insulators
with narrow valence bands in many ionic crystals, as well as
Ne,Ar,Kr,Xe solids (assuming the lattice constant fixed.).
Furthermore, it is not difficult to see that the GGA or the
X$_\alpha$ method becomes invalid in calculating the ionization
potential for the extended, including noncrystalline, systems, where
the valence electrons cannot be described by the plane waves. As we
have already observed, the GGA or the X$_\alpha$ method would
predict that the hole state is extended over many different regions
which are distant from each other even in inhomogeneous systems.
Such prediction differs from the commonsense.

 In section III, we have also
pointed out that the magnitude of the total electronic energy is
significantly overestimated by the GGA in the large $a$ limit when
the band is not completely filled.  Actually, Zhang and Yang and
other groups \cite{SIE1,SIE2,SIE3} also found that H$_2^+$,F$_2^+$,
NaCl et.al. would not be dissociated correctly in the GGA in the
large distance limit. This deficiency is more pronounced in solids
because the electronic states are more extended. It should be
emphasized that the self-interaction of the electronic states in our
system is negligibly small.  Therefore, this deficiency cannot be
removed by the so called self-interaction-corrected local density
approximation method\cite{SIC1,SIC2}.

 In summary, we speculate that the exact
exchange-correlation functional is highly nonlocal when the electron
density $n({\bf r})$ varies in space and the system is extended in
the density functional theory\cite{DFT}. When the density of the
valence electrons varies greatly in space and the system is
extended, the GGA and the X$_\alpha$ method may be unsuitable to
describe the system properly.


\begin{thebibliography}{Zhang 97}

\bibitem{DFT} P. Hohenberg, W. Kohn, {\it Phys. Rev.} {\bf 136},
B864(1964).

\bibitem{LDA} W. Kohn, L.J. Sham, {\it Phys. Rev.} {\bf 140}, A1133(1965).

\bibitem{GGA1} J.P. Perdew, {\it Phys. Rev. Lett.} {\bf 55}, 1665(1985).

\bibitem{GGA2} J.P. Perdew and Wang Yue, {\it Phys. Rev.}{\bf B33},
8800(1986).


\bibitem{Xa} J.C. Slater, {\it Phys. Rev.} {\bf 81}, 385(1951).

\bibitem{Xa23} R. Gaspar, {\it Acta Phys. Acad. Sci. Hung.}, {\bf
3},263(1954).

\bibitem{Levine} See for instance, I.N. Levine, $\ll${\it Quantum
Chemistry}$\gg$ Fifth Edition, Prentice Hall, Upper Saddle River,
New Jersey(2000).

\bibitem{Kunz} R. Pandey, J.E. Jaffe, and A.B. Kunz, {\it Phys.
Rev.} {\bf B43}, 9228(1991).

\bibitem{units} All numerical numbers appeared in this paper are
given in atomic units.


\bibitem{SIE1} Y. Zhang, W. Yang, {\it J. Chem. Phys.}{\bf
109},2604(1998).

\bibitem{SIE2} R. Merble, A. Savin, and H. Preuss, {\it J. Chem.
Phys.}{\bf 97}, 9216(1992).

\bibitem{SIE3} T. Bally, G.N. Sastry, {\it J. Phys. Chem.}{\bf
A101},7923(1997).

\bibitem{SIC1}  J.P. Perdew and A. Zunger, {\it Phys. Rev.}{\bf B23},
5048(1981).

\bibitem{SIC2} S. Goedecker and C. Umrigar, {\it Phys. Rev.}{\bf
A55}, 1765(1997).


\end{thebibliography}
\end{document}